\documentstyle[12pt]{article}
\begin{document}

\title{Density Contrast--Peculiar Velocity Relation in the Newtonian Gauge }

\author{Sohrab Rahvar\footnote{E-mail:rahvar@theory.ipm.ac.ir}\\
Department of Physics, Sharif University of Technology,\\
P.O.Box 11365--9161, Tehran, Iran\\
and\\
Institute for Studies in Theoretical Physics and Mathematics,\\
P.O.Box 19395--5531, Tehran, Iran}\maketitle

\begin{abstract}
In general relativistic framework of large scale structure
formation theory in the universe, we investigate relation between
density contrast and peculiar velocity in the Newtonian gauge.
According to the gauge--invariant property of the
energy--momentum tensor in the this gauge, the velocity
perturbation behaves as the Newtonian peculiar velocity. In this
framework, relation between peculiar velocity and density
contrast with respect to the Newtonian Peebles formula has an
extra correction term which is ignorable for the small scales
structures. The relativistic correction of peculiar velocity for
the structures with the extension of hundred mega parsec is about
few percent which is smaller than the accuracy of the recent
peculiar velocity measurements. We also study CMB anisotropies
due to the Doppler effect in the Newtonian gauge comparing with
using the Newtonian gravity.
\end{abstract}
\maketitle
\section{Introduction}

In the large structure formation theory, structures originate from
small density fluctuations that are amplified by gravitational
instability. These initial fluctuations are assumed to be
generated during inflationary epoch and are inflated to the
beyond of the horizon while the size of horizon remains constant
during the inflation. These quantum fluctuations beyond the
horizon freeze and evolve like classical perturbations. They
reenter the horizon at a later time when the horizon grows to the
size of perturbation. The observational evidence for these small
density fluctuations can be seen at the decoupling epoch as the
anisotropy of CMB and also existence of the large scale structures
in the universe.\\
One of the consequences of density fluctuation is metric
perturbations in the Friedman--Robertson--Walker (FRW) universe
and deviation of velocity field from the Hubble law which is
called {\it peculiar velocity}. Hence studying peculiar velocity
could be a useful indirect method to find out the structures in
the FRW universe. In the Newtonian linear structure formation
theory, relation between density contrast and peculiar velocity
has been obtained by Peebles \cite{pee80}. From the experimental
point of view, Bertschinger and Dekel introduced POTENT method to
reconstruct velocity potential from the radial component of
velocity field \cite{bert89}. Recently Branchini {\it et al}
combined measurement of radial velocity of galaxies and
independent measurement of density contrast to
estimate $\Omega$ \cite{dek00}.\\
On the scales well below than the Hubble radius, the Newtonian
theory of gravitation is a good approximation and Peebles'
approach is widely used. However for observations and simulations
are probing scales which are a significant fraction of the Hubble
radius, the light-cone effect should taken into account
\cite{man02}. Our aim in this work is to generalize relation
between density contrast and peculiar
velocity to general relativistic framework. \\
The relativistic linear perturbation of FRW universe can be
studied by two type of formulations exist in the literature. The
'gauge fixing' formulation, originated by Lifshitz \cite{lif46},
considers perturbed components of the metric which are related to
the energy momentum tensor. According to the gauge freedom
problem in the theory, it is  necessary to put a constraint on
perturbed part of metric for gauge fixing (e.g. synchronous
gauge). In the Second approach, initiated by Bardeen in 1980
\cite{bar80} gauge invariant variables can be made by the
combining perturbed metric elements. These gauge-invariant
variables are well known from other physical theories. For
example, in classical electrodynamics it is usually more physical
to work in terms of gauge-invariant electromagnetic fields rather
than in terms of gauge-dependent scalar and vector potentials. The
subsequent works in FRW perturbation theory can
be found in some textbooks \cite{wei72}-\cite{ber00}. \\
General relativistic analysis of peculiar velocity has been
studied in the 'covariant' fluid flow \cite{bru94}, 'Harmonic
gauge' \cite{man02} and  ' quasi-Newtonian gauge' fixing methods.
In this work we are going to use conformal Newtonian gauge also
known as the (Longitudinal gauge) to obtain relation between
density contrast and peculiar velocity. The conformal Newtonian
gauge  was introduced by Mukhanov {\it et al} \cite{muk92} as a
simple gauge used for scalar mode of metric perturbation. In this
gauge perturbation of metric elements are gauge-invariant
variables. For the case of absence of non-diagonal space--space
component in energy--momentum tensor, perturbation of metric can
be interpreted as relativistic generalization of Newtonian
gravitational potential. We generalize this idea for peculiar
velocity field according to its gauge--invariant property in the
Energy--Momentum tensor. Here, we restrict ourselves to the flat
universe, in the matter and radiation dominant epochs, and obtain
relativistic relation between density contrast and peculiar
velocity field. It is shown that in small scales compare to the
size of horizon, where general relativistic effect due to the
light cone effect is negligible, our formulation reduces to
Peebles equation. For the structures with $300Mpc$ extension, the
relativistic correction is about one percent. We also calculate
the relativistic Doppler effect contribution on the CMB
anisotropies and compare it with the Newtonian one. \\
The organization of the paper is as follows. In Sec. II we write
down perturbation of metric in Newtonian gauge. In Sec. III we
introduce energy--momentum tensor and Einstein equations in this
gauge and In Sec. IV, we obtain relativistic relation between
density contrast and peculiar velocity for flat universe in
radiation and matter dominant epochs. In Sec. V , we obtain
signature of Doppler effect on the anisotropies of CMB, using
relativistic peculiar velocity. We conclude in Sec. VI with a
brief summary and some discussions.

\section{Perturbation of FRW in the Newtonian Gauge}

Consider a small perturbation of metric $h_{\mu\nu}$ with respect
to FRW  background $g_{\mu\nu}$
\begin{equation}
\label{metric}
 ds^2 = (g_{\mu\nu} + h_{\mu\nu})dx^{\mu}dx^{\nu}
\end{equation}
where $g_{\mu\nu} = {a(\tau)}^2(d\tau^2 - \gamma_{ij}dx^idx^j)$
and $\tau$ is conformal time. Greek and Roman letters go from $0$
to $3$ and $1$ to $3$, respectively. The metric perturbation
$h_{\mu\nu}$ could be categorized into three distinct types like
Scalar, vector and tensor perturbation. This classification
refers to the way in which $h_{\mu\nu}$ transforms under
three--space coordinate transformation on the constant time
hyper-surface.
A covariant description of tensor decomposition has been shown in
\cite{ste90}. Vector and tensor modes exhibit no instability. Vector
perturbation decays kinematically in an expanding universe and tensor
perturbation leads to
gravitational waves which do not couple to energy density and
pressure inhomogeneities. However, scalar perturbation may lead to
growing inhomogeneities of matter. \\
The most general form of the scalar metric perturbation is
constructed by using four scalar quantities $\phi,\psi, B$ and
$E$:
\label{h}
\begin{equation}
h_{\mu\nu}^{(s)}  =  \: \left(
\begin{array}{cc}
2\phi & -B_{;i}\\
B_{;i} & 2(\psi\gamma_{ij} -E_{;ij})
\end{array} \; \; \right)\;.
\end{equation}
One of the main difficulties of the relativistic structure
formation theory is that there is gauge freedom in the theory and
one can make an artificial perturbation by coordinate
transformation in such a way that infinitesimal space--time
distance between the two event remains constant. One way to
overcome this problem is using gauge fixing and the other way is
gauge invariant method \cite{bar80}. In the latter method, some
quantities are found to be gauge invariant under coordinate
transformation. It seems that the gauge--invariant quantities are
similar to the electric and magnetic fields in the theory of
electrodynamics where they are measurable quantities in contrast
to the vector and scalar potentials that are gauge dependent
parameters. The simplest gauge--invariant quantities from the
linear combination of $\phi, \psi, B$ and $E$ which span the
two--dimension space of gauge--invariant variables are:
\begin{eqnarray}
\Phi^{(gi)} &=& \phi + \frac{1}{a}[(B-{E}^{'})a]^{'} \\
\Psi^{(gi)} &=& \psi - \frac{{a}^{'}}{a}(B -{E}^{'})
\end{eqnarray}
where prime represents derivation with respect to conformal time
and index {\it (gi)} stands for gauge invariance. The above
variables were first introduced by Bardeen \cite{bar80}. Newtonian
gauge is defined by choosing $B= E =0$ \cite{muk92}. It is seen
that in this gauge $\phi$ and $\psi$ become gauge--invariant
variables and the metric can be written as follows:
\begin{equation}
\label{metric}
 ds^2 =
{a(\tau)}^2\left[(1+2\phi)d\tau^2-(1-2\psi)\gamma_{ij}dx^idx^j\right].
\end{equation}

\section{Einstein Equation in The Newtonian Gauge}

In the perturbed background of the FRW metric, the linear
perturbation of Einstein equation can be written as follows:
\begin{equation}
\label{perturb}
 \delta G^{\mu}{}_{\nu} = 8\pi G\delta T^{\mu}{}_{\nu}.
\end{equation}
The left hand side of eq.(\ref{perturb}) can be obtained from
perturbation of metric in the Newtonian gauge and the right hand
side of it is obtained by perturbed energy--momentum tensor.
Restricting our attention to scalar perturbation of
energy--momentum tensor, we can express the most general first
order energy--momentum tensor in terms of four scalars as follows
\cite{bar80}:
\begin{equation}
\label{energy}
 \delta T^{\mu}{}_{\nu}  =  \: \left(
\begin{array}{cc}
\delta\rho & -(\rho + p)a^{-1}u_i\\
(\rho + p)au_i & -\delta p + \sigma_{;ij}
\end{array} \; \; \right)\; ,
\end{equation}
where $\delta\rho$ and $\delta p$ are the perturbations of energy
density and pressure, ${\cal W}$ is potential for the
three--velocity field $v^i = a(t) u^{i}(x,\tau)=\nabla^i {\cal W}$
and $\sigma$ is anisotropic stress. Bardeen has shown that in the
Newtonian gauge: $\delta T^{\mu}{}_{\nu}^{(gi)} = \delta
T^{\mu}{}_{\nu}$ \cite{bar80}. According to this result, $v^i
=a(t)u^i$ may be regarded as Newtonian peculiar velocity. For a
perfect fluid, the anisotropic stress which leads to non-diagonal
space--space component of the energy--momentum tensor vanishes.
In this case it can be shown that $\phi=\psi$. Using the metric
of eq.(\ref{metric}) in the left hand side and eq.(\ref{energy})
in the right hand side of eq.(\ref{perturb}), Einstein equation
can be rewritten:
\begin{eqnarray}
\label{cont} & & \nabla^{2}\phi - 3{\cal{H}}{\phi}' -
3({\cal{H}}^2 -
\kappa)\phi = 4\phi G\delta\rho, \\
\label{u}
& & {(a\phi)}'_{,i} = -4\phi G (\rho + p)u_{i},\\
\label{p} & & {\phi}'' + 3{\cal H}{\phi}' + (2{\cal H}^{'} +{\cal
H}^2 -\kappa )\phi = 4\phi G \delta p ,
\end{eqnarray}
where ${\cal H}$ is the Hubble constant in conformal time. For
simplicity in calculation we consider the following change of
variable.
\begin{equation}
\label{phi} \phi = 4\pi G(\rho+p)^{1/2} \omega = (4\pi
G)^{1/2}[({\cal H}^2 - {\cal H}^{'} + \kappa)a^2]^{1/2}\omega.
\end{equation}
By substituting eq.(\ref{phi}) into eq.(\ref{p}), this equation is
rewritten in terms of $\omega$ as follows:
\begin{equation}
\label{om} {\omega}^{''} - {c_s}^2\nabla^2\omega
-\frac{{\theta}^{''}}{\theta}\omega = 0,
\end{equation}
where
\begin{equation}
\label{tetha} \theta = \frac1a(\frac{\rho}{\rho +
p})^{1/2}(1-\frac{3\kappa}{8\pi G\rho
a^2})^{1/2}.
\end{equation}
 and for the adiabatic perturbations $c_s^2 = \frac{\delta p}{\delta\rho}$. For the case of
 matter dominant epoch one can obtain exact solution for differential equation
(\ref{om}), considering $(c_s=0)$, yields:
\begin{equation}
\label{omeg}
\omega(x,\tau) = E_1(x)\theta(\tau)
+E_2(x)\theta(\tau)\int\frac{d\tau}{\theta^2}.
\end{equation}
Our aim is to obtain density contrast and peculiar velocity in
terms of $\omega$. Density contrast $\delta =
\frac{\delta\rho}{\delta}$ can be obtained by dividing
eq.(\ref{cont}) to $G^{0}{}_{0}$ component of the FRW equation
$({\cal H}^9 + \kappa = \frac{8\pi G \rho a^2}{3}$):
\begin{equation}
\label{delta}
 \delta = \frac{2}{3({\cal H}^2
+\kappa)}(\nabla^2\phi - 3{\cal H}\phi^{'}-3({\cal H}^2 -
\kappa)\phi).
\end{equation}
 In what follows, Eqs.(\ref{u}) and (\ref{delta}) will be our main equations in the
following sections. It is seen that these two equations are as
function of $\phi$ which, can be obtained by solving
eq.(\ref{om}). In the next section we find an explicit relation
between density contrast and peculiar velocity in the framework
of general relativistic perturbation theory in the Newtonian
gauge. we restrict ourselves to the case of radiation and matter
dominant epochs in the flat universe.
\section{Density Contrast--Peculiar Velocity Relation}
From the recent CMB and Supernova type I experiments, it seems
that the curvature of the universe is flat \cite{lan00} ($\kappa
= 0)$. In this section according to sequence of the radiation and
matter dominant epochs from early universe, we apply
Eqs.(\ref{u}) and (\ref{delta}) for these two regimes.

\subsection{ Radiation dominant epoch}
In the radiation dominant epoch, for the flat universe, according
to FRW equations, scale factor evolves like $a = a_0\tau$. In
this case eq.(\ref{tetha}) reduce to:
\begin{equation}
\label{tetha3}
 \theta = \sqrt{\frac 34}\frac 1a .
\end{equation}
For simplicity in the calculation we consider structures larger
than Hubble radius, so the second term in the right hand side of
eq.(\ref{om}) can be ignored with respect to other terms and
$\omega(x,\tau)$ can be obtained form eq.(\ref{omeg}).
Substituting eq.(\ref{tetha3}) into eq.(\ref{omeg}), the dynamics
of $\omega$ is obtained:
\begin{equation}
\label{om2} \omega = \frac{1}{({\cal H}^2 - {\cal
        H}^{'})^{4/2}}(\frac{D(x,\tau)}{a})^{'}  ,
\end{equation}
where $D(x,\tau) = [E_1 \sin(\nu\tau)
+E_2\cos(\nu\tau)]e^{(ik\cdot x)}$ and $\nu =
\frac{k}{\sqrt{3}}$. Substituting eq.(\ref{om2}) into
eqs.(\ref{phi}), (\ref{delta}) and (\ref{u}), we obtain following
equations for $\phi(x,\tau)$, $\delta(x,\tau)$ and $u_i(x,\tau)$:
\begin{eqnarray}
\phi(x,\tau)&=& \frac{1}{{\tau}^3}\left[(\nu\tau\cos(\nu\tau) -
                \sin(\nu\tau))C_1 - (\nu\tau\sin(\nu\tau) +
                \cos(\nu\tau))C_6\right]e^{(ik\cdot x)}  ,\\
\label{d1}
 \beta(x,\tau)&=& \frac {4}{{\tau}^3}[([(\nu\tau)^2 -
                1]\sin(\nu\tau) +
                \nu\tau[1 -
                \frac12(\nu\tau)^2]\cos(\nu\tau)C_1 \nonumber\\
                &+&([1-(\nu\tau)^2]\cos(\nu\tau) + \nu\tau[1 -
                \frac12(\nu\tau)^2]\sin(\nu\tau))C_2)]e^{(ik\cdot
                x)}  ,\\
\label{vv}
 \nabla\cdot v &=&
                 \frac{a_1 k^2}{a^2 \tau}(\left[(\nu\tau\cos(\nu\tau) -
                 \cos(\nu\tau))C_1 - (\nu\tau\sin(\nu\tau) +
                  \cos(\nu\tau))C_2\right]e^{(ik\cdot x)} \nonumber\\
                  &-&\frac{k^2a_0}{2a^2}[(\nu\cos(\nu\tau)-
                  \nu^2\tau\sin(\nu\tau) -
                  \nu\cos(\nu\tau))C_1\nonumber\\
                  &-&(\nu\sin(\nu\tau)+\nu^2\tau\cos(\nu\tau)
                   -\nu\sin(\nu\tau))C_2])e^{(ik\cdot x)}  .
\end{eqnarray}
According to our consideration for the structures larger than
Hubble radius $\nu\tau<<1$, eqs.(\ref{d1}) and (\ref{vv}) are
reduced as follows:
\begin{eqnarray}
\label{v2}
\nabla\cdot v &=& -\frac{k^2a_0}{a^5\tau}e^{(ik\cdot x)}C_2,\\
\label{de2} \delta &=& \frac{4}{\tau^3}e^{(ik\cdot x)}C_2.
\end{eqnarray}
Now one can divide eq.(\ref{v2}) by eq.(\ref{de2}) to obtain the
explicit relation between density contrast and the peculiar
velocity:
\begin{equation}
\label{v3}
 \nabla\cdot v =
-\frac34 H\delta({\nu\tau})^2.
\end{equation}
It is seen from eq.(\ref{v3}) that in the radiation dominant
epoch, the relativistic peculiar velocity for a given density
contrast is much smaller than what one can expect from the
Newtonian formula.
\subsection{Matter dominant epoch}
 In the matter dominant epoch, for $\kappa = 0$,
eq.(\ref{tetha}) is simplified into this form:
\begin{equation}
\label{the1}
 \theta = \frac{1}{a} .
\end{equation}
For the flat universe, the scale factor grows as $a =
a_{0}\tau^2/2$. By substituting eq.(\ref{the1}) in
eq.(\ref{omeg}), the dynamics of $\omega$ as a function of
conformal time is obtained:
\begin{equation}
\omega = E_1(x)\tau^3 + E_2(x)\tau^{-2} ,
\end{equation}
where $E_1(x)$ and $E_2(x)$ are arbitrary functions of spatial
coordinates. Using $\omega$ in the eq.(\ref{phi}), the value of
$\phi$ obtain as follows:
\begin{equation}
\label{phi1} \phi(x,\tau) = C_1(x) + \tau^{-7}C_2(x) .
\end{equation}
Substituting eq.(\ref{phi1}) into eqs.(\ref{u}) and
(\ref{delta}), $\delta$ and $u$ are obtained as functions of
conformal time, $C_1(x)$ and $C_2(x)$ as:
\begin{eqnarray}
\label{del1}
\delta &=& \frac16\left[(\nabla^2C_6(x) - 81C_2(x)) -
\tau^{-5}(\tau^2\nabla^2C_2(x) + 18C_2(x))\right],\\
\label{u1}
u_i(x,\tau) & = & \frac1a_5(\frac{C_2(x)_{,i}}{\tau^6} -
\frac23\frac{C_1(x)_{,i}}{\tau}).
\end{eqnarray}
Neglecting the decaying modes, we rewrite eqs.(\ref{del1}) and
(\ref{u1}) in the Fourier space:
\begin{eqnarray}
\label{del2}
\delta_k &=&-\frac16(k^2\tau^2 +12)C_1(k),\\
\label{u2}
\frac{ik \cdot v_i(k)}{a} & =&
\frac{2}{3a_1}\frac{k^2}{\tau}C_1(k).
\end{eqnarray}
Dividing eq.(\ref{u2}) by eq.(\ref{del2}), one can obtain an
explicit relation between density contrast and peculiar velocity
in the following form:
\begin{equation}
\label{kv}
 \frac{ik.v_k}{a} = - \frac{H\delta_k}{1 + \frac{3a^2H^2}{k^2}}
\end{equation}
where, H is the Hubble constant in physical time and $v_k$ is the Fourier
transformation of the peculiar velocity. By using the definition of
scalar potential for the peculiar velocity, $v_k = \frac{ik}{a}{\cal
W}_{k}$, eq.(\ref{kv}) can be written:
\begin{equation}
\label{de}
 H\delta_k = \frac{k^2}{a^2}{\cal W}_{k} + 3H^2{\cal W}_{k}.
\end{equation}
In the real space, eq.(\ref{de}) changes to:
\begin{equation}
\label{vv1} \nabla\cdot v = - H\frac{\delta}{b} +3H^2 {\cal W}
\end{equation}
where $b$ is biasing factor and ${\cal W}= \int v \cdot dl$ is the
scalar potential of velocity field and can be obtained by radial
integrating of velocity field. The second term in the right hand
side of eq.(\ref{vv1}) is a result of general relativistic
corrections. This correction can be estimated by dividing the
second term by the first term in the right hand side of
eq.(\ref{de}).
\begin{equation}
\label{relat} \frac{3H^2W_k}{\frac{k^2}{a^2}W_k}
\simeq(\frac{\lambda}{d_H})^2
\end{equation}
where $d_H$ is the size of horizon and $\lambda = \frac{a}{k}$ is
the size of structure. Is is seen that for the small scale
structures ( compare to the size of horizon), the general
relativistic correction is negligible. This correction for the
structures with $300Mpc$ extension is about one percent.
\section{The Effect of Relativistic Peculiar Velocity on CMB}
The inhomogeneities in the universe induce anisotropies in the
distribution of relic background of the photons on the CMB. The
anisotropy of CMB can be caused by several reasons. One of the
main reasons is that the matter which scattered the radiation in
our direction had a peculiar velocity with respect to comoving
frame when the scattering occurred. Since the peculiar velocity of
matter is different at different directions in the sky, this leads
to an anisotropy on the CMB. The universe at the last scattering
epoch has a redshift of $z \approx 1000$. Since the equality of
matter and radiation occurred at $z\approx 5\times10^4$, we can
consider that at last scattering epoch universe resides in the
matter dominant. Using eq.(\ref{kv}) for relativistic peculiar
velocity in matter dominant epoch, the relativistic anisotropy
due to Doppler effect on CMB is obtained as follows:
\begin{equation}
\label{T} \frac{\delta T}{T} \simeq v
=\frac{\frac{\lambda}{d_H}}{1+3(\frac{\lambda}{d_H})^2}\delta
\end{equation}
where $\lambda$ is the size of structure and $d_H$ is the Hubble
radius at the decoupling epoch. It is seen from eq.(\ref{T}) that
the temperature perturbation of the CMB due to peculiar
velocity in the relativistic framework can be expressed in
the term of Newtonian one:
\begin{equation}
\label{nn}
 (\frac{\delta T}{T})_{Rel} = \frac{(\frac{\delta
T}{T})_{New}}{1+3(\frac{\theta}{\theta_H})^2}
\end{equation}
where $(\frac{\delta T}{T})_{New} = \frac{\lambda}{d_H}\delta$ is
the contribution of Newtonian Doppler anisotropy, $\theta =
34^{''}(\Omega h)(\frac{\lambda}{3Mpc})$ is the angular size of
structure \cite{pad93} and $\theta_H \simeq 1^\circ$ is the
angular size of Hubble radius.  It is seen from eq.(\ref{nn})
that for the large angles, the relativistic contribution of
Doppler effect due to the peculiar velocity on the fluctuations
of CMB is smaller than the contribution of
Newtonian anisotropy. \\
We compare the relativistic Doppler effect also with Sachs--Wolf
effect (which is one of the reasons for the the anisotropy of CMB
arising from the variation in the gravitational potential at the
last scattering \cite{sac67}). It is shown that the contribution
of this effect on anisotropy of CMB can be given by:
\begin{equation}
\label{schcs} \frac{\delta T}{T} =
\frac{1}{3}(\frac{\theta}{\theta_H})^2 \delta.
\end{equation}
Eqs.(\ref{T}) and (\ref{schcs}) shows that Sachs--wolf effect at
large angles $(\theta>\theta_H)$ is dominant term on the
anisotropies of CMB while for the small angles Doppler term is
dominant. Fig. 1 shows the fluctuation of temperature as a
function of angular size of the anisotropy for the case of
Newtonian and relativistic peculiar velocity and Sachs--wolf
effect which normalized to the density contrast. It can be shown
that the contribution of relativistic peculiar velocity affect
the power spectrum of CMB.
\section{Conclusion}
In this letter, we have tried to identify the peculiar velocity in
general relativistic theory of structure formation. We obtain a
relation between the density contrast and the peculiar velocity
for the matter and radiation dominant epochs. It has been shown
that the relativistic correction is about few percent for
structures with extension of hundred mega parsec. The effect of
the relativistic peculiar velocity as Doppler effect on the CMB
also has been calculated. It was shown that the contribution of
relativistic peculiar velocity on the anisotropy of CMB is less
than what one expected from the Newtonian theory.

\newpage

\section{Figure Caption}
This figure shows the normalized perturbation of temperature to
density contrast $(\frac{\delta T}{T})/ \delta$ as a function of
normalized angular size of structures to the size of horizon
$\frac{\theta}{\theta_H}$. Solid line, dashed line and
dashed--dot line represent the contribution of Newtonian,
Relativistic Doppler effect and Sachs--Wolf effect on the
anisotropy of CMB, respectively.


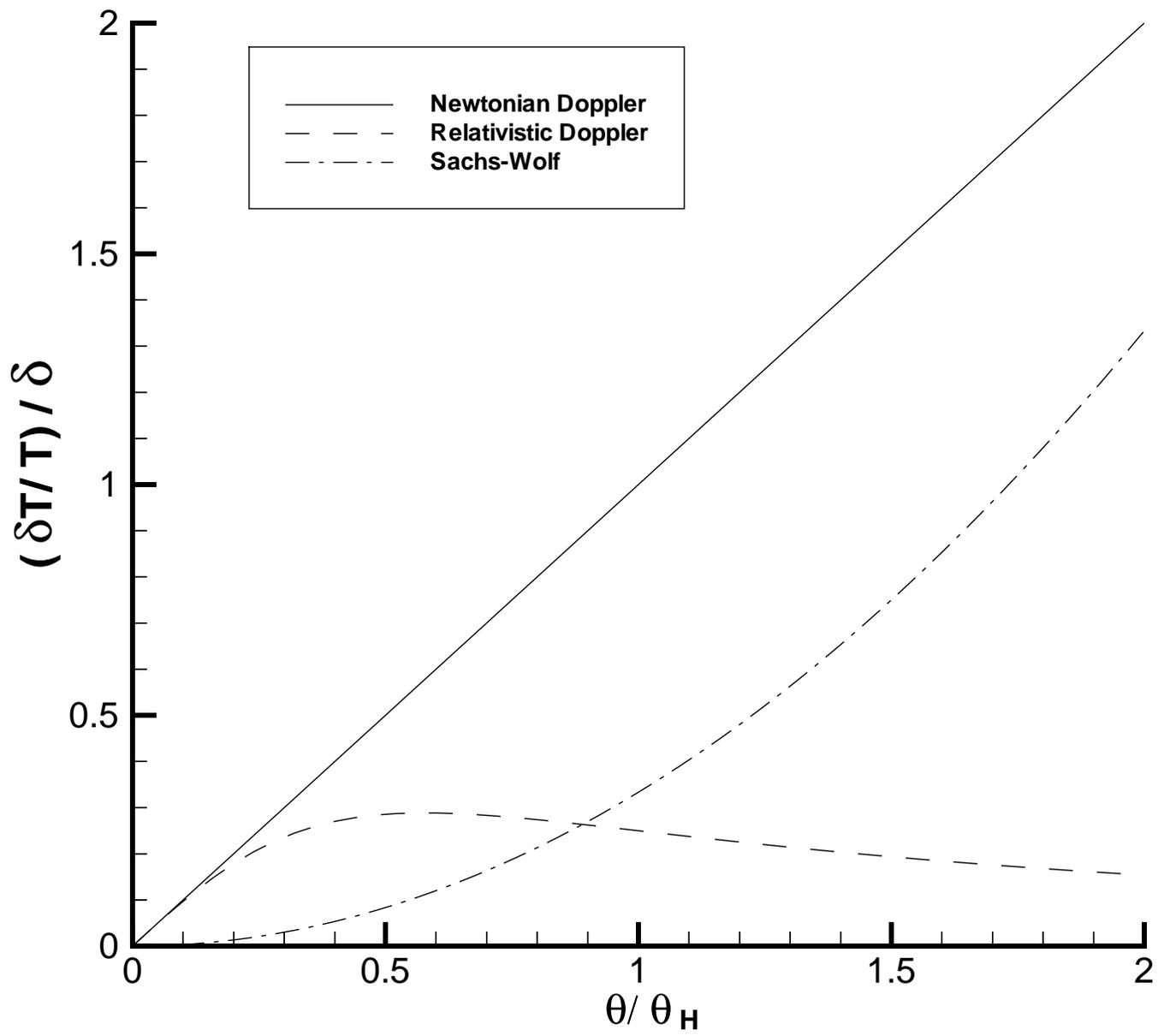
\end{document}